\documentclass[preprint]{aastex}
\usepackage{natbib}
\usepackage{booktabs}
\usepackage{threeparttable}

\usepackage{longtable}
\usepackage{tabularx}
\usepackage{lineno}

\newcommand{\fexxi}{Fe \scriptsize{XXI} \normalsize}
\newcommand{\fexviii}{Fe \scriptsize{XVIII} \normalsize}
\newcommand{\fei}{Fe \scriptsize{I} \normalsize}

\newcommand{\ci}{C \scriptsize{I} \normalsize}
\newcommand{\oi}{O \scriptsize{I} \normalsize}
\newcommand{\siiv}{Si \scriptsize{IV} \normalsize}

\begin{document}

\title{Observational signatures of electron-driven chromospheric evaporation in a white-light flare}

\author{Dong~Li\altaffilmark{1,2}, Chuan~Li\altaffilmark{3,4}, Ye~Qiu\altaffilmark{3,4}, Shihao~Rao\altaffilmark{3,4}, Alexander~Warmuth\altaffilmark{5}, Frederic~Schuller\altaffilmark{5}, Haisheng~Zhao\altaffilmark{6}, Fanpeng~Shi\altaffilmark{1,7}, Jun~Xu\altaffilmark{1,7}, and Zongjun~Ning\altaffilmark{1,7}}

\affil{$^1$Key Laboratory of Dark Matter and Space Astronomy, Purple Mountain Observatory, CAS, Nanjing 210023, PR China \\
       $^2$State Key Laboratory of Space Weather, Chinese Academy of Sciences, Beijing 100190, PR China \\
     $^3$School of Astronomy and Space Science, Nanjing University, Nanjing 210023, PR China \\
     $^4$Key Laboratory of Modern Astronomy and Astrophysics (Nanjing University), Ministry of Education, Nanjing 210023, PR China \\
     $^5$Leibniz-Institut f\"{u}r Astrophysik Potsdam (AIP), An der Sternwarte 16, Potsdam 14482, Germany \\
     $^6$Key Laboratory of Particle Astrophysics, Institute of High Energy Physics, CAS, Beijing 100049, PR China \\
     $^7$School of Astronomy and Space Science, University of Science and Technology of China, Hefei, 230026, PR China}
     \altaffiltext{}{Correspondence should be sent to: lidong@pmo.ac.cn}

\begin{abstract}
We investigate observational signatures of explosive chromospheric
evaporation during a white-light flare (WLF) that occurred on 2022
August 27. Using the moment analysis, bisector techniques,
and the Gaussian fitting method, red-shifted velocities of less
than 20~km~s$^{-1}$ are detected in low-temperature spectral lines of
H$\alpha$, \ci\ and \siiv\ at the conjugated flare kernels, which
could be regarded as downflows caused by chromospheric condensation.
Blue-shifted velocities of $\sim$30$-$40~km~s$^{-1}$ are found in
the high-temperature line of \fexxi, which can be interpreted as
upflows driven by chromospheric evaporation. A nonthermal hard X-ray
(HXR) source is co-spatial with one of the flare kernels, and the
Doppler velocities are temporally correlated with the HXR fluxes. The
nonthermal energy flux is estimated to be at least
(1.3$\pm$0.2)$\times$10$^{10}$~erg~s$^{-1}$~cm$^{-2}$. The
radiation enhancement at \fei~6569.2~{\AA} and 6173~{\AA} suggests
that the flare is a WLF. Moreover, the while-light emission at
\fei~6569.2~{\AA} is temporally and spatially correlated with the
blue shift of \fexxi\ line, suggesting that both the white-light
enhancement and the chromospheric evaporation are triggered and
driven by nonthermal electrons. All our observations support the
scenario of an electron-driven explosive chromospheric evaporation
in the WLF.
\end{abstract}


\section{Introduction}
Solar flares are detected as dramatic and impulsive electromagnetic
emissions over a wide wavelength range on the Sun. In flares, energy
which has been stored in nonpotential coronal magnetic field is
released suddenly and converted to thermal energy of hot plasmas,
kinetic energy of nonthermal particles, and bulk mass motions
\citep{Shibata11,Benz17,Jiang21}. The energy release is presumably
caused by magnetic reconnection, as described by the two-dimensional
(2-D) standard model \citep{Priest02,Lin05} or the CSHKP flare model
\citep{Carmichael64,Sturrock66,Hirayama74,Kopp76}. In the vicinity
of the reconnection region, the local plasma escapes in the form of
bi-directional outflow jets, largely due to the strong curvature of
reconnected magnetic field lines
\citep[e.g.,][]{Mann11,Lid19,Warmuth20,Li21,Yan22}. They are usually
turbulent and may contain plasmoids. The outflowing plasma
is heated, and electrons are accelerated to nonthermal energies by a
yet undetermined mechanism. Both heat conduction and accelerated
electrons can provide a high energy flux into the dense
chromosphere, which reacts with explosive heating and expansion.
This results in hot plasma flowing up and quickly filling the newly
formed flare loops, a process which is known as ``chromospheric
evaporation'' \citep{Fisher85,Brosius15,Dudik16,Yang21}. Conversely,
plasma can also  slowly move downward due to momentum conservation,
which is termed as ``chromospheric condensation''
\citep{Kamio05,Libbrecht19,Graham20,Yu20}.

Chromospheric evaporation exists in two variants: explosive
and gentle. The two forms are sensitive to the incident
energy flux, i.e., a threshold of 10$^{10}$~erg~s$^{-1}$~cm$^{-2}$
\citep[e.g.,][]{Fisher85,Kleint16,Sadykov19}. It should be
pointed out that the incident energy threshold will also depend on
the location of energy deposition and the duration of heating
\citep{Reep15,Polito18}, as well as the spectral parameters of the
electrons (low-energy cutoff and slope). Explosive evaporation may
also be driven by nonthermal electrons accelerated by a stochastic
process at a low energy flux, i.e.,
$<$10$^{10}$~erg~s$^{-1}$~cm$^{-2}$ \citep{Rubio15}. In the case of
explosive chromospheric evaporation, chromospheric plasmas are
rapidly heated, resulting in a much higher pressure in the
chromosphere. Then, some plasmas expand upward into the low-density
corona at fast velocities of several hundreds km~s$^{-1}$, while
some other plasmas precipitate downward with a slow velocity of a
few tens km~s$^{-1}$ into the high-density chromosphere due to
momentum conservation
\citep[e.g.,][]{Veronig10,Brosius17,Tian18,Sellers22}. That
is, the upflows driven by chromospheric evaporation and the
downflows caused by chromospheric condensation can be simultaneously
observed in an explosive evaporation event. In contrast, only the
upflows can be seen for gentle evaporation. It is difficult to
observe the downflows in gentle evaporation, which could be
attributed to a much higher plasma density in the lower chromosphere
\citep{Doschek13}. Line profile spectroscopy is a good tool for
diagnosing the chromospheric evaporation and condensation,
which are often identified as the upflows of
high-temperature plasmas in the corona and the downflows of
low-temperature plasmas in the chromosphere, respectively
\citep[e.g.,][]{Teriaca12,Doschek13,Tian14,Tian21,Polito17,Dep21}.
In spectroscopic observations
\citep[e.g.,][]{Ding96,Milligan09,Li15,Lee17,Brosius18,Chen22}, the
upflow is often identified as the blue-shifted Doppler
velocity of hot lines at coronal heights, i.e., \fexviii, \fexxi,
etc., and the downflow is often regarded as the red shift of
cool lines in the chromosphere or transition region, such as
H$\alpha$, \ci, \siiv, etc. Therefore, the gentle evaporation is
characterized by blue shifts in all spectral lines
\citep{Milligan06a,Li19,Sadykov19}, while the explosive type is
characterized by blue shifts in high-temperature lines and red
shifts in low-temperature lines
\citep{Milligan06b,Zhang16,Brosius17}.

Chromospheric evaporation can also be seen in soft X-ray (SXR) and
Extreme Ultra-Violet (EUV) images \citep{Liu06,Ning10,Ning11}, or
observed in the radio dynamic spectrum \citep{Aschwanden95,Ning09}.
The rapid mass motions originate from double footpoints along the
flare loop and subsequently merge together at the loop-apex region.
The moving speed could be equal to the upflow speed of hot
evaporated materials. This phenomenon is regarded as the imaging
evidence of chromospheric evaporation \citep[e.g.,][]{Zhang19,Li22}.
The higher frequency suddenly suppresses and drifts to the lower
frequency is regarded as the radio signature of chromospheric
evaporation \citep{Aschwanden95}. The chromospheric evaporation is
frequently detected in the solar flare
\citep{Chen10,Young15,Polito17,Libbrecht19}, and it can be observed
during the precursor phase \citep{Brosius10,Li18}, the impulsive
phase \citep{Graham15,Li17,Sadykov19} or the gradual phase
\citep{Czaykowska99,Brannon16}. The similar evaporation phenomenon
could also be found in the solar flare-like transient
\citep{Brosius07} or the coronal bright point \citep{Zhang13,Lid22},
because all these solar eruptive events are basically characterized
by hot plasmas generated by the magnetic reconnection. Chromospheric
evaporation can be driven by the nonthermal electrons
\citep{Tian15,Warren16,Lee17}, the thermal conduction
\citep{Battaglia15,Ashfield21}, or the dissipation of Alfv\'{e}n
waves \citep{Reep16,Van20}. The electron-driven model emphasizes
that nonthermal energies play a key role during the chromospheric
evaporation, while the thermal conduction driven model states that
thermal energies can directly cause the chromospheric evaporation.

A white-light flare (WLF) is a type of solar flare that
shows a sudden enhancement in the white-light continuum, which was
first reported by \cite{Carrington59} and \cite{Hodgson59}.
White-light emission is often thought to be formed in the lower
atmosphere on the Sun, but the generation mechanism of white-light
emission is still unclear \citep{Aboudarham89,Fang13}. Several
heating mechanisms have been discussed in the literature, for
instance nonthermal electron bombardment \citep{Hudson72}, thermal
irradiation \citep{Henoux77}, and chromospheric condensation
\citep{Gan94}. Recently, magnetic reconnection in the lower solar
atmosphere has been proposed to power a WLF \citep{Song20}. In this
paper, we explore observational signatures of explosive
chromospheric evaporation during a WLF, which are consistent with
the electron-driven model. Section~2 presents instruments and
observations. Section~3 describes data analyses and results.
Section~4 offers a brief discussion, and Section~5 summarizes our
conclusion.

\section{Instruments and Observations}
The solar flare studied in this work occurred on 2022
August 27 in the active region of NOAA 13088 (S26W66). It was
simultaneously measured by the Chinese H$\alpha$ Solar
Explorer\footnote{https://ssdc.nju.edu.cn} \citep[CHASE;][]{LiC22},
the Interface Region Imaging Spectrograph \citep[IRIS;][]{Dep14},
the Atmospheric Imaging Assembly \citep[AIA;][]{Lemen12} and the
Helioseismic and Magnetic Imager \citep[HMI;][]{Schou12} aboard the
Solar Dynamics Observatory (SDO), the Spectrometer/Telescope for
Imaging X-rays \citep[STIX;][]{Krucker20} on Solar Orbiter
\citep{Muller20}, the Gravitational wave high-energy Electromagnetic
Counterpart All-sky Monitor \citep[GECAM;][]{Xiao22}, the
Geostationary Operational Environmental Satellite (GOES), and the
STEREO/WAVES (SWAVES), as listed in table~\ref{tab1}.

CHASE is designed to acquire the spectroscopic observations at
wavebands of H$\alpha$ and \fei. The scientific payload is the
H$\alpha$ Imaging Spectrograph (HIS), which has two observational
modes: raster scanning mode (RSM) and continuum imaging mode (CIM).
Here, the RSM data products are used for analysis, and they
provide full-Sun spectral images at wavelength ranges of
6559.7~{\AA}$-$6565.9~{\AA} (H$\alpha$) and
6567.8~{\AA}$-$6570.6~{\AA} (\fei). In normal observation, the
spectral dispersion is 24~m{\AA} pixel$^{-1}$, and the pixel scale
is 0.52\arcsec\ \citep{Qiu22}. For the in-orbit binning data applied
in this study, the spectral and spatial resolutions are
twice lower than nominal. The time cadence is about 71~s.
IRIS is a space-based multi-channel UV imaging spectrograph
mainly focused on the solar chromosphere and transition region
\citep{Dep14}. On 2022 August 27, IRIS observed the active region in
a large ``sit-and-stare'' mode with a time cadence of
$\sim$9.4~s. The pixel scale along the slit is $\sim$0.166\arcsec.
Here, we use the spectral lines at IRIS windows of
``\oi~1356~{\AA}'' and ``\siiv~1403~{\AA}'', and the Slit-Jaw Imager
(SJI) at 1400~{\AA}. The far ultraviolet (FUV) spectra at ``\oi''
and ``\siiv'' had a spectral dispersion of
25.96/25.44~m{\AA}~pixel$^{-1}$. SJI~1400~{\AA} images had a
field-of-view (FOV) of 119\arcsec$\times$119\arcsec\ with a time
cadence of 28.2~s. SDO/AIA is designed to take full-disk solar
images in multiple wavelengths nearly simultaneously
\citep{Lemen12}. SDO/HMI is designed to investigate the solar
magnetic field and oscillatory features on the Sun \citep{Schou12}.
Here, the AIA images at seven EUV channels of 94~{\AA}, 131~{\AA},
171~{\AA}, 193~{\AA}, 211~{\AA}, 304~{\AA}, and 335~{\AA} are used,
which have a time cadence of 12~s. We also use the continuum
filtergram and the line-of-sight (LOS) magnetogram of SDO/HMI, which
have a time cadence of 45~s. The AIA and HMI maps have been
calibrated by using the standard programs of aia\_prep.pro and
hmi\_prep.pro, respectively, and thus they all have a spatial scale
of 0.6\arcsec~pixel$^{-1}$.

STIX provides X-ray imaging spectroscopy of solar flares over an
energy range of 4$-$150~keV and thus characterizes both the hot and
super-hot flare plasmas as well as the accelerated nonthermal
electrons \citep{Krucker20}. In this study, we use two different
STIX data products. The spectrogram data, which is generated
on-board by summing over all pixels in all detectors, is used to
generate HXR countrates in wider energy bins, such as 4$-$10~keV,
10$-$15~keV, and 15$-$25~keV. The advantage of this data type is
that we can use the full temporal resolution of STIX, which was
0.5~s in this event. In contrast, the pixelated science data is
obtained by downlinking the counts for all pixels in all detectors
individually. We use this data set to generate disk-integrated
spectrograms and to reconstruct HXR images. For this event, the time
resolution of the pixelated data was set at 20~s in order to ensure
statistically significant numbers of counts in all pixels. We wanted
to state that all STIX times have been modified in order to be
consistent with the observations from 1~AU, as Solar Orbiter was at
0.79~AU in this event.

GECAM is designed to monitor high-energy astrophysical features,
i.e., searching for gravitational waves, detecting $\gamma$-ray
bursts. It consists of 25 Gamma Ray Detectors (GRD) and 8 Charged
Particle Detectors (CPD). GECAM/GRD can also provide the solar
irradiance measurement in hard X-rays and $\gamma$-rays
\citep{Xiao22}. In this study, the solar HXR flux at the high gain
channel such as 20$-$60~keV is used for analysis, which has an
uniform time cadence of 0.2~s.

\section{Analyses and Results}
\subsection{Overview of the solar flare}
\begin{figure}
\centering
\includegraphics[width=\linewidth,clip=]{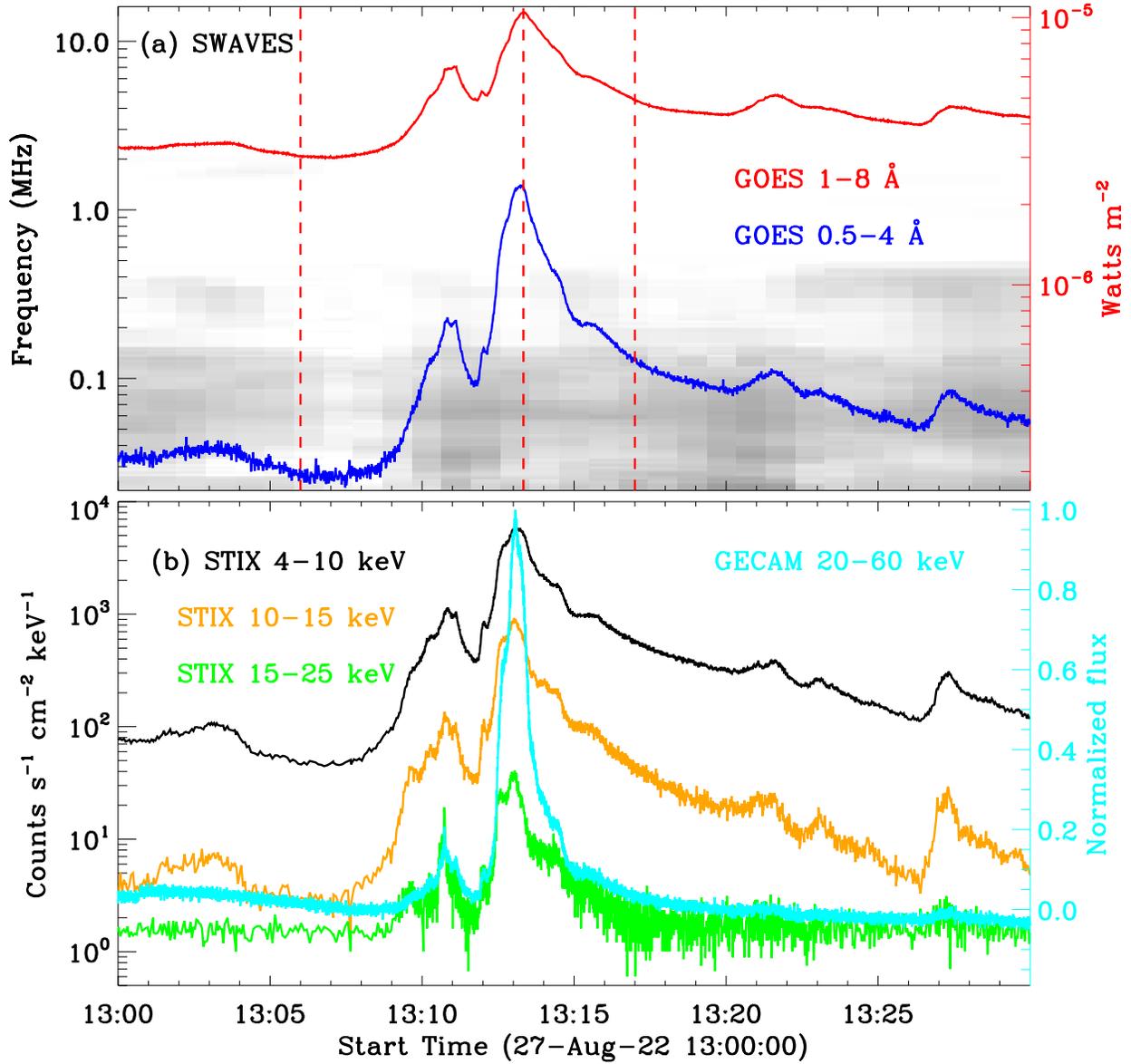}
\caption{Full-disk light curves from 13:00~UT to 13:30~UT on 2022
August 27. Panel~(a): GOES SXR fluxes at wavelengths of 1$-$8~{\AA}
(red) and 0.5$-$4~{\AA} (blue). The context image is the radio
spectrogram observed by SWAVES. Panel~(b): STIX count fluxes in the
energy bands of 4$-$10~keV (black), 10$-$15~keV (orange), and
15$-$25~keV (green), and the normalized GECAM flux at 20$-$60~keV
(cyan). \label{flux}}
\end{figure}

Figure~\ref{flux} shows an overview of X-ray light curves integrated
over the entire Sun during 13:00$-$13:30~UT on 2022 August 27.
Panel~(a) presents SXR fluxes recorded by GOES in 1$-$8~{\AA} (red)
and 0.5$-$4~{\AA} channel (blue). It shows a C9.6 class
flare that begins at about 13:06~UT, reaches its maximum at around
13:13~UT, and stops at 13:17~UT, as indicated by the red vertical
lines. Both GOES fluxes reveal two main peaks, implying two episodes
of energy releases. The background image is the radio dynamic
spectrum measured by SWAVES at lower frequencies between 0.023~MHz
and 16.025~MHz. The radio spectrum does not show any signature of
solar radio bursts, suggesting that there are not nonthermal
electrons escaping from the flare region. The temporal
resolution of SWAVES is one minute, but it still can resolve the
presence of type III radio bursts at lower frequencies. Moreover, we
do not see any signature of type III radio bursts in solar dynamic
spectra of
e-Callisto\footnote{http://soleil.i4ds.ch/solarradio/callistoQuicklooks/}
with high temporal resolution (i.e., $\leq$1~s). No coronal mass
ejection (CME)\footnote{https://www.sidc.be/cactus/catalog.php} was
detected after the C9.6 flare. All those facts suggest that it is a
confined flare. Figure~\ref{flux}~(b) plots the X-ray light curves
measured by STIX at 4$-$10~keV (black), 10$-$15~keV (orange), and
15$-$25~keV (green), and GECAM~20$-$60~keV (cyan), respectively. It
can be seen that all the light curves show two main pulses,
similarly to what has observed by GOES. Moreover, the second pulse
is much stronger than the first one, and we fucus on the second HXR
pulse, which spans from the impulsive phase to the gradual phase of
the C9.6 flare.

\begin{figure}
\centering
\includegraphics[width=\linewidth,clip=]{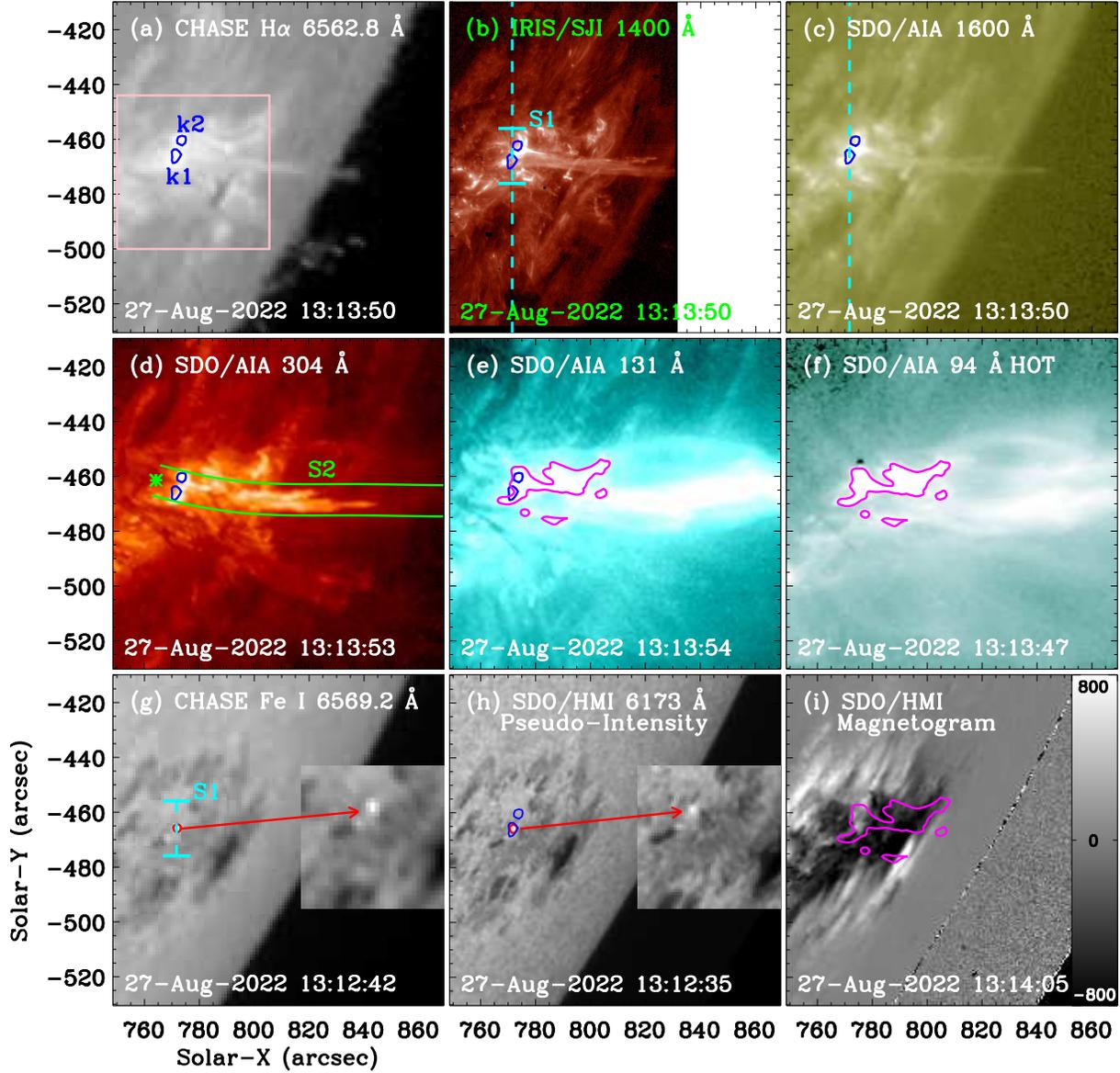}
\caption{Snapshots in multiple wavebands with a same FOV of
120\arcsec$\times$120\arcsec\ measured by CHASE/HIS (a, g), IRIS/SJI
(c), SDO/AIA (c$-$f), and SDO/HMI (h, i). Blue and magenta contours
outline bright kernels/features in H$\alpha$ and AIA~94~{\AA} hot
channels. The pink box and two short cyan lines mark the flare
region, and the dashed cyan line (S1) represents the slit of IRIS.
The green (S2) curves outline the solar jet, which are used to
perform the time-distance maps in Figure~\ref{slice2}. The red arrow
indicates an enhancement of the white light radiation. Two zoomed
images show \fei~6569.2~{\AA} and HMI pseudo-intensity maps with a
small FOV of about 30\arcsec$\times$30\arcsec. \label{imag}}
\end{figure}

In Figure~\ref{imag}, we present multi-wavelength maps with a same
FOV of 120\arcsec$\times$120\arcsec. Panel~(a) shows the H$\alpha$
map at 6562.8~{\AA} during the flare observed by CHASE, which
reveals two bright flare kernels (k1 and k2), as outlined by the
blue contours. They could be regarded as the conjugated footpoints
of a hot flare loop. Panels~(b) plots the FUV image captured by
IRIS/SJI in 1400~{\AA}, while panels~(c) and (d) draw UV/EUV images
observed by SDO/AIA at wavelengths of 1600~{\AA} and 304~{\AA}. They
all exhibit two bright and compact kernels, which match well with
double H$\alpha$ kernels, as indicated by the blue contours.
Therefore, these maps at wavelengths of H$\alpha$~6562.8~{\AA},
SJI~1400~{\AA}, and AIA~1600~{\AA} are used to co-align between
different instruments. The dashed cyan line marks the slit of IRIS,
which goes through one flare kernel (i.e., k1). A solar jet
originated from the loop-apex appear simultaneously in those four
images, and it is most vivid in AIA~304~{\AA}, as outlined by two
green lines.

Figures~\ref{imag}(e) and \ref{imag}(f) show EUV maps at
high-temperature channels of AIA~131~{\AA} and 94~{\AA}. A short
flare loop that connects to the double H$\alpha$ kernels (blue
contours) can be seen in AIA~131~{\AA}. It is well known that the
two AIA channels in 131~{\AA} and 94~{\AA} are sensitive to hot
plasmas, but a strong contribution from cool plasmas is also found
in these two channels \citep[e.g.,][]{ODwyer10}. Therefore,
we utilize the approach following the Equation~\ref{eq1} to extract
the hot plasma from AIA~94~{\AA} \citep[cf.][]{Del13,Gupta18}.

\begin{equation}
  I_{\rm hot} \approx I({\rm 94~{\AA}})-\frac{I({\rm 171~{\AA}})}{450}-\frac{I({\rm 211~{\AA}})}{120},
\label{eq1}
\end{equation}
\noindent Where, I({\rm 94~{\AA}}), I({\rm 171~{\AA}}), and I({\rm
211~{\AA}}) represent intensities measured by SDO/AIA at the
corresponding wavelengths. Panel~(f) shows the EUV map at
AIA~94~{\AA} hot channel, and the short flare loop can be seen at
the high-temperature channel, as indicated by the magenta contour.
Moreover, it links to the conjugated footpoints that root in
positive and negative magnetic fields, respectively, as shown in
panel~(i).

Figures~\ref{imag}(g) and \ref{imag}(h) present white-light
images measured by CHASE and SDO/HMI during the C9.6 flare,
respectively. The CHASE map at \fei 6569.2~{\AA} reveals one bright
feature, as outlined by the red contour. This bright feature could
be considered as the enhancement of white-light radiation, and it
can be also seen in the pseudo-intensity map, as shown in the two
zoomed images. We should state that the pseudo-intensity map
is made from SDO/HMI continuum filtergrams at two adjacent times,
since it is easier to observe the white-light enhancement even when
it is rather weak \citep[cf.][]{Song18,Li23}. The formation height
of \fei\ line at 6569.2~{\AA} is estimated to be in the
mid-photosphere around 250~km for the quiet Sun
\citep[cf.][]{Hong22}. Therefore, the C9.6 flare is indeed a
white-light flare. Moreover, the white-light enhancement locates in
one H$\alpha$ kernel that scanned by the IRIS slit, as indicated by
the dashed cyan line. Here, we also generate the time-distance
images along the IRIS slit (short dashed cyan line) that derived
from the CHASE maps at \fei~6569.2~{\AA} and H$\alpha$~6562.8~{\AA}.

\subsection{CHASE spectra}
\begin{figure}
\centering
\includegraphics[width=\linewidth,clip=]{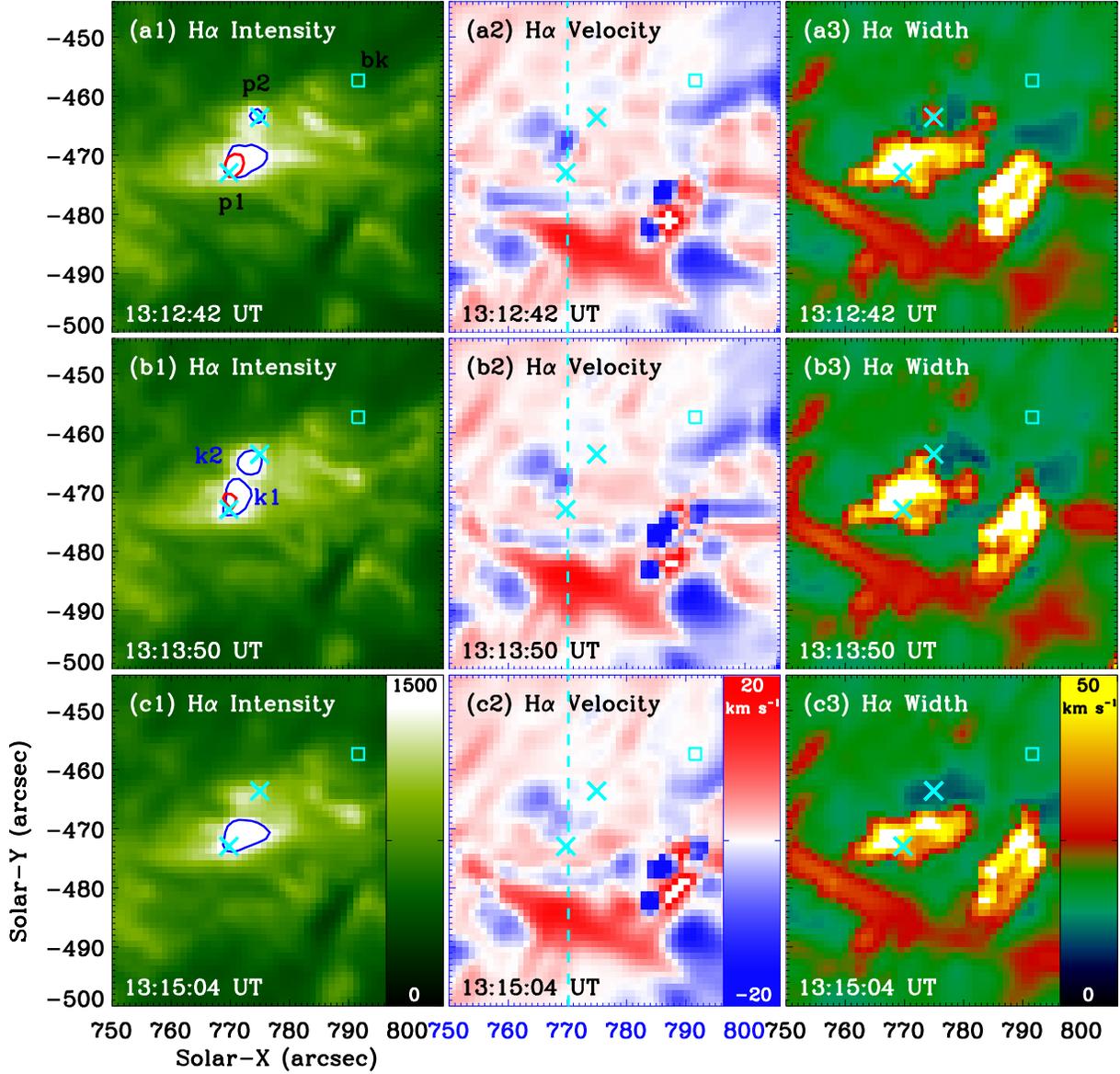}
\caption{Moment analysis results (line intensity, Doppler velocity,
and line width) from CHASE H$\alpha$ spectra during the C9.6 flare.
They all have a small FOV of 56\arcsec$\times$56\arcsec, as outlined
by the pink box in Figure~\ref{imag}~(a). Blue and red contours
represent the H$\alpha$ line intensity and \fei\ emission,
respectively. Two cyan crosses (`$\times$') and one box mark the
locations to show H$\alpha$ spectra at flare kernels and non-flare
(background) region in Figure~\ref{sp_ha2}. The dashed cyan line
represents the slit of IRIS. \label{sp_ha1}}
\end{figure}

Two flare kernels (k1 and k2) can be seen in the CHASE H$\alpha$ map
at 6562.8~{\AA}, and the IRIS slit covers one of them (i.e., k1), as
shown in Figure~\ref{imag}. The H$\alpha$ spectrum, which is
recorded by the CHASE full-Sun raster scanning, contains the
spectral and spatial information of the Sun and can be used to
investigate the double flare kernels. Based on the moment analysis
technique \citep[e.g.,][]{Brannon16,Liy17,Li20}, the 2-D maps of
line intensity, Doppler velocity and width at H$\alpha$ waveband can
be obtained, which are determined from the zeroth, first, and second
moments \citep[cf.][]{Yu20}. Figure~\ref{sp_ha1} shows H$\alpha$
maps for the line intensity (left), Doppler velocity (middle), and
line width (right) at three times during the C9.6 flare. They all
have a FOV of about 56\arcsec$\times$56\arcsec, as outlined by the
pink box in Figure~\ref{imag}~(a).

Similarly to what one sees in the H$\alpha$ map at 6562.8~{\AA}, the
line intensity maps basically reveal two flare kernels, as shown in
Figure~\ref{sp_ha1}~(a1)$-$(c1). The k2 kernel appears and enhances
from 13:12:42~UT to 13:13:50~UT, then it becomes very weak and
gradually disappears at about 13:15:04~UT. At the same
time, the k1 kernel can always be seen from 13:12:42~UT to
13:15:04~UT during the C9.6 flare, and it is much stronger than the
k2 kernel. The white-light emission at CHASE \fei\ shows one bright
source, which is co-spatial with the H$\alpha$ kernel k1, as
indicated by the red contour in panel~(a1). However, the white-light
emission only lasts for a short time, and becomes weak at
13:13:50~UT and gradually disappears at 13:15:04~UT, as shown in
panels~(b1) and (c1). The Doppler velocities of these two kernels
are dominated by red shifts, as shown in
Figure~\ref{sp_ha1}~(a2)$-$(c2), and their Doppler velocities
are small, i.e., $<$10~km~s$^{-1}$. Their line widths are
also enhanced at two flare kernels, and the line width at k1 kernel
is significantly stronger than that at k2 kernel. The feature of
Doppler velocities and line widths implies that the chromospheric
condensation is simultaneously observed at two flare kernels. We
also note that the red-shifted velocity tends to appear at the outer
edge of flare kernels, which is similar to previous observations
\citep{Czaykowska99,Li04}

\begin{figure}
\centering
\includegraphics[width=\linewidth,clip=]{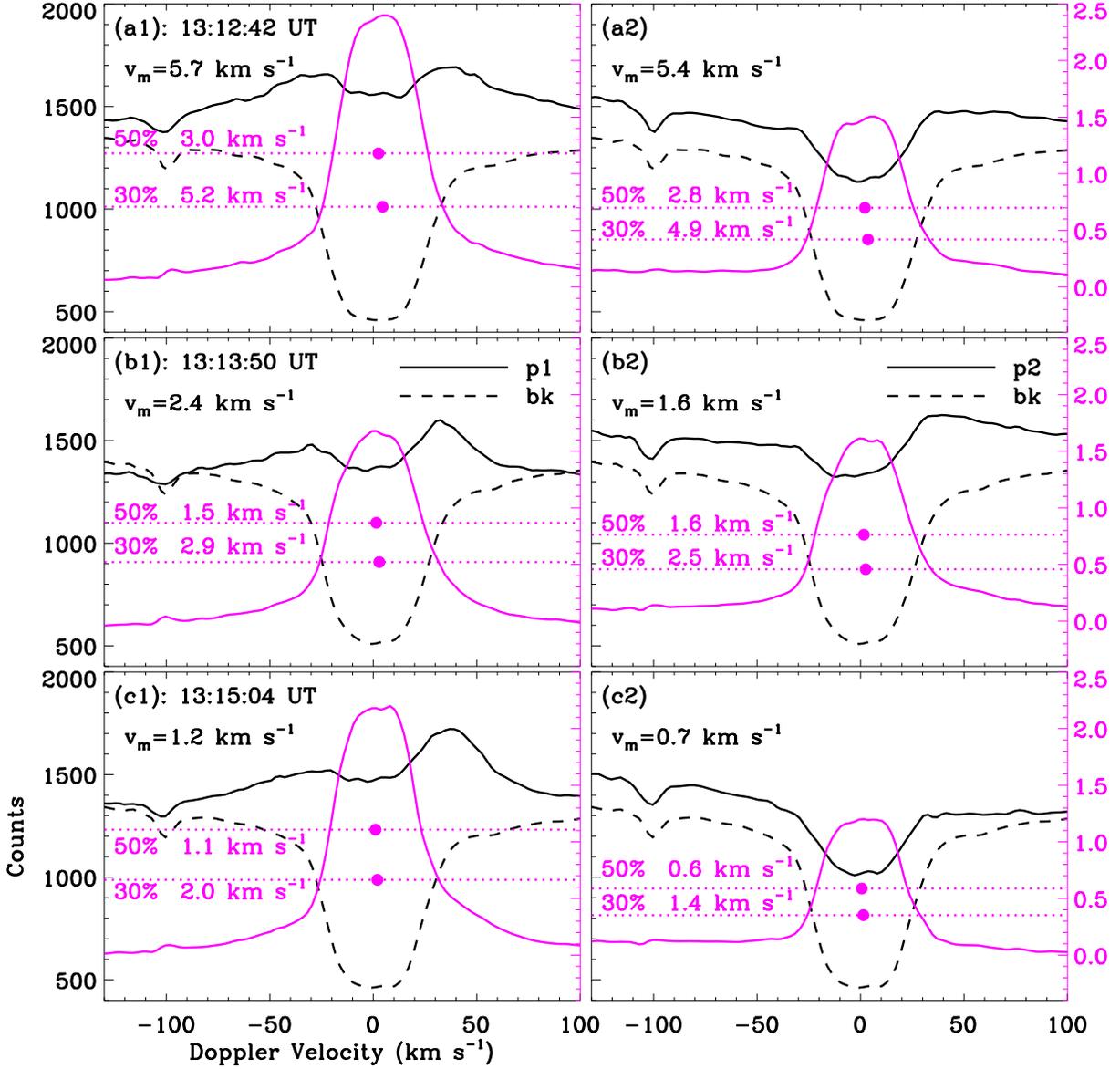}
\caption{H$\alpha$ spectra observed by CHASE/HIS in flare kernels
(solid black) and non-flare region (dashed black) at three time. The
solid magenta curves indicate the H$\alpha$ contrast line profiles,
as normalized by the background spectrum. The dotted magenta lines
represent emission levels, and the bisector velocities are indicated
by solid dots. \label{sp_ha2}}
\end{figure}

In order to look closely at the H$\alpha$ spectrum,
Figure~\ref{sp_ha2} presents the line profiles of H$\alpha$ at the
locations of two kernels and the background (no-flare region), as
marked by the crosses and box in Figure~\ref{sp_ha1}. The black
solid and dashed lines represent H$\alpha$ line profiles at flare
kernels and background, respectively. Similar to previous
observations \citep[e.g.,][]{Ding94,Li04}, the H$\alpha$ line cores
at flare kernels are central-reversed. The H$\alpha$ line
profile at the background seems to be symmetrical. At the
same time, those line profiles at flare kernels appear to be
extremely asymmetrical, and the enhancement at the red-wing is
stronger than that at the blue-wing, confirming the red-shifted
velocities at flare kernels. To measure their Doppler shifts, we
perform the moment analysis technique \citep[e.g.,][]{Liy17,Yu20} to
the H$\alpha$ contrast line profiles, which are derived from the
Equation~\ref{eq2}:
\begin{equation}
  C (\lambda) = (I_{\rm k}(\lambda)-I_{\rm b}(\lambda))/I_{\rm b}(\lambda),
\label{eq2}
\end{equation}
\noindent Where $C (\lambda)$ represents the contrast H$\alpha$ line
profile as a function of the wavelength ($\lambda$), $I_{\rm
k}(\lambda)$ and $I_{\rm b}(\lambda)$ are H$\alpha$ line profiles at
the flare kernel and background, respectively. The magenta curves in
Figure~\ref{sp_ha2} are H$\alpha$ contrast line profiles at flare
kernels, and they all demonstrate red shifts of less than
10~km~s$^{-1}$. The smallest velocity is found at the k2 kernel at
13:15:04~UT (c2), and the line intensity and width are also
very weak. This is consistent with the observation fact that the
H$\alpha$ emission here is very weak.

As is known, the moment method only provides averaged parameters. In
order to further confirm the Doppler velocities at two H$\alpha$
kernel positions, we also perform the bisector method to the
emission components of H$\alpha$ contrast line profiles
\citep[cf.][]{Ding95,Hong14,Ishikawa20}. In this method, the Doppler
velocity is calculated from the shift of the line center relative to
the rest wavelength, and the line-center shift is determined as the
middle point's wavelength of the H$\alpha$ contrast line profile
($C$) at a given emission level ($q$), such as
$q*(C_{max}-C_{min})+C_{min}$. Here, $C_{max}$ and $C_{min}$
represent the maximal and minimal values of the H$\alpha$ contrast
emission. The dotted magenta lines in Figure~\ref{sp_ha2} indicate
50\% and 30\% emission levels, and their bisector velocities are
also marked by the magenta dots. Obviously, the bisector velocities
are sensitive to emission levels, but they all reveal red shifts at
slow velocities, i.e., $<$10~km~s$^{-1}$, which are comparable to
the moment velocities ($v_m$). Thus, we can conclude that the two
H$\alpha$ kernels show red shifts, and they could be regarded as
downflows caused by chromospheric condensation.

\subsection{IRIS spectra}
\begin{figure}
\centering
\includegraphics[width=\linewidth,clip=]{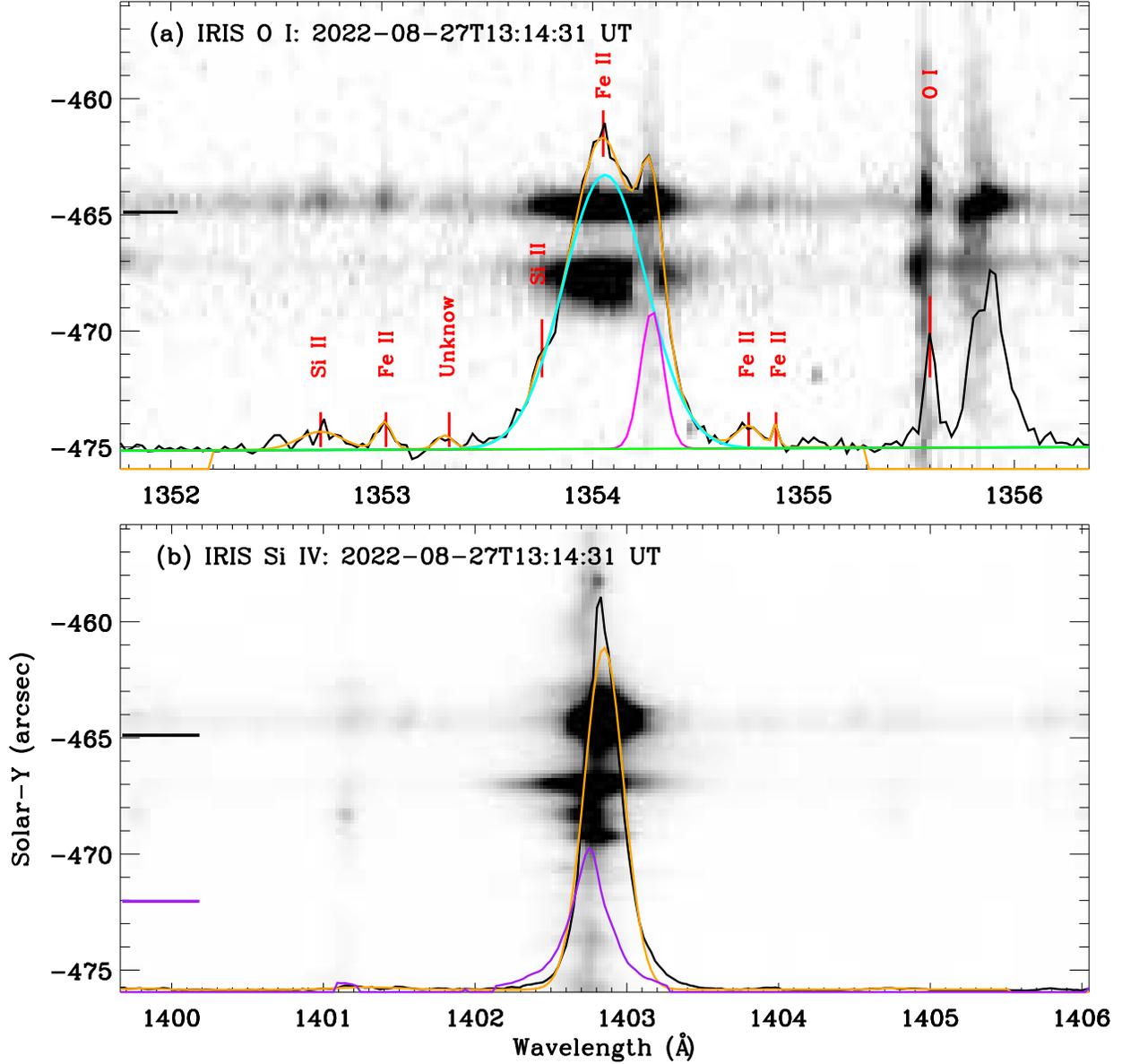}
\caption{IRIS spectra at two FUV windows of `\oi' (a) and `\siiv'
(b) on 2022 August 27. The black profiles are line spectra at the
flare kernel, as marked by the short black line on the left side of
each panel, and the overplotted orange profiles represent their
best-fitting results. The cyan and magenta lines are \fexxi\ and
\ci, and the green line is the fitting background. The red vertical
ticks labeled some other emission lines used in this study. The
purple curve (after multiplication by 10) is the line spectrum at
non-flare location, as marked by the short purple line.
\label{sp_iris1}}
\end{figure}

In order to study the flare kernel (i.e., k1) in details, we then
analyse IRIS spectral lines, including \fexxi, \ci, and \siiv. The
forbidden transition of \fexxi\ line is a potentially useful
diagnostic for high-temperature ($\sim$11~MK) plasmas during the
solar flare, while the chromospheric line of \ci\ can be
used for the diagnostics of low-temperature ($\sim$0.01~MK)
plasmas. However, these two and some other emission lines in
the considered spectral window are blended
\citep[e.g.,][]{Tian15,Young15,Li16,Polito16,Tian17}. Thus, we
utilize multi-Gaussian function superimposed on a linear background
to fit the IRIS spectrum \citep[cf.][]{Li15,Li17,Li18}.
Figure~\ref{sp_iris1}~(a) presents the IRIS spectrum in the window
of ``\oi'' during the C9.6 flare. It can be seen that a broad line
is blended with some other emission lines, as indicated by the short
vertical red lines. Here the emission line \oi is applied for
absolute wavelength calibration \citep{Tian15}. The orange line
represents the best-fitting results by using the multi-Gaussian
function, which matches well with the observed line spectrum. The
green line is the linear background, while the cyan and magenta
lines indicate the line profiles of \fexxi\ and \ci, respectively.
In the IRIS window of ``\siiv'', the transition line of \siiv\ is
isolated, and it is much stronger than other emission lines during
the solar flare, as shown in Figure~\ref{sp_iris1}~(b).
Thus, we use a single-Gaussian function (orange) to fit the
line spectrum (black). The reference wavelength of \siiv\ is
determined from the non-flare region, as indicated by the purple
line. On the other hand, the reference wavelength of \fexxi\ cannot
be determined from the non-flare region, because it is a hot flare
line. So, the reference wavelength of \fexxi\ is selected as
1354.1~{\AA} \citep[cf.][]{Young15}. Then the Doppler
velocity of each emission line can be determined by subtracting the
fitted line center from its reference wavelength.

\begin{figure}
\centering
\includegraphics[width=\linewidth,clip=]{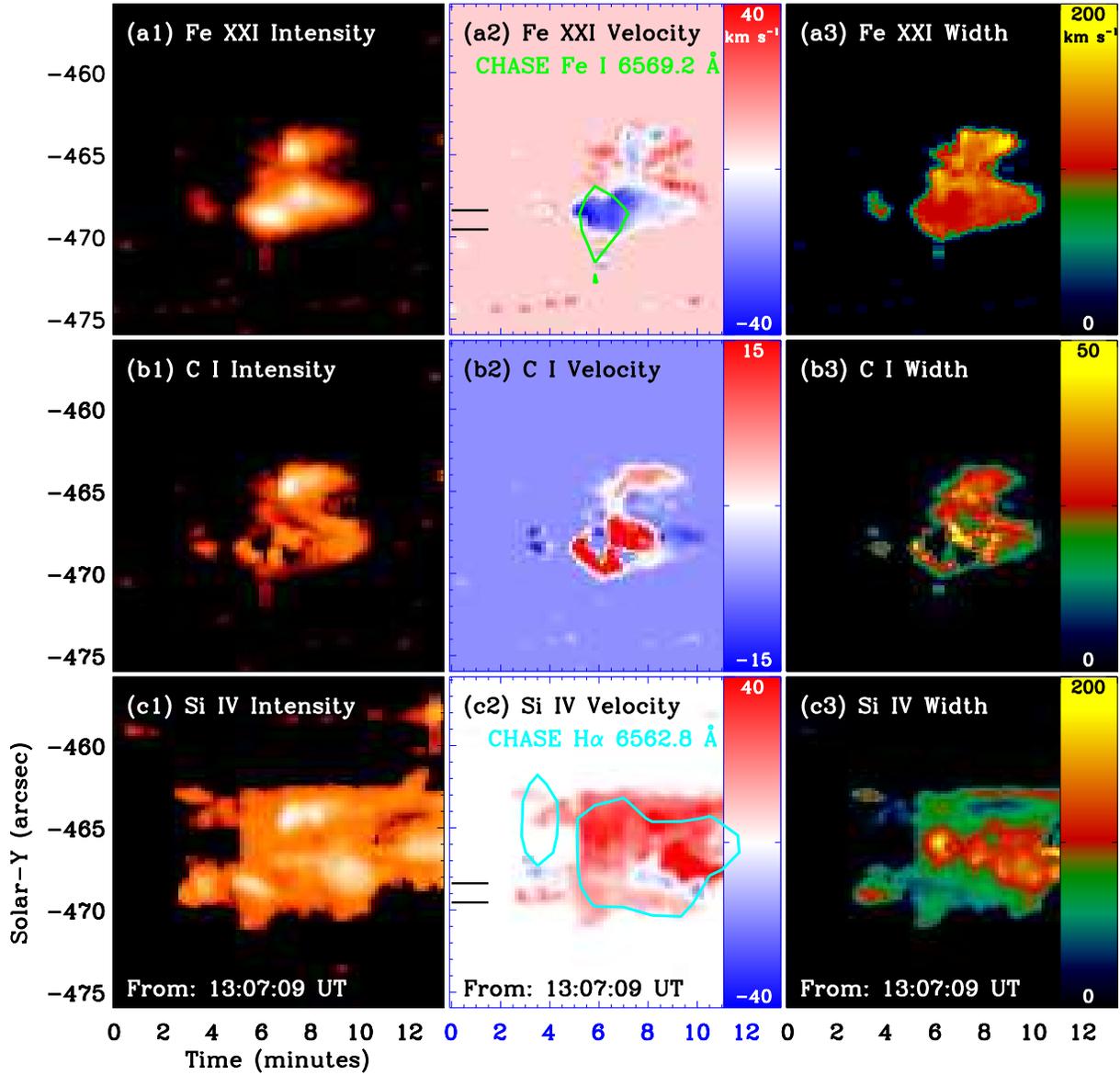}
\caption{Time-distance images of line intensity, Doppler velocity
and line width from IRIS spectra, e.g., \fexxi, \ci, and \siiv. Two
short black lines outline the flare region used to integrate the
time series in Figure \ref{corr}. Green and cyan contours represent
the CHASE observations at \fei\ and H$\alpha$. A logarithmic scale
is used for the line intensity. \label{sp_iris2}}
\end{figure}

Figure~\ref{sp_iris2} shows time-distance diagrams of line
intensity, Doppler velocity and line width for \fexxi, \ci, and
\siiv, respectively. The displayed images have the
same length of about 20\arcsec\ along the IRIS slit, as indicated by
two short cyan line in Figure~\ref{imag}~(b). From the left panels,
we notice that a very weak FUV emission appears in the flare kernel
during the first HXR pulse, i.e., $\sim$13:09$-$13:11~UT. Then, a
strong FUV emission is seen at these three emission lines, which
corresponds to the second HXR pulse. During the second HXR pulse,
the Doppler velocity of \fexxi\ line tends to show blue shifts of
$\sim$30$-$40~km~s$^{-1}$, while that of \ci\ and \siiv\
lines appears to show red shifts of
$\sim$10$-$20~km~s$^{-1}$, as shown in the middle panels. Their line
widths are all broad at flare kernels, as shown in the right panels.
We note that the speeds of hot evaporating plasmas measured
from \fexxi line blueshifts are pretty weak, which is in
disagreement with the expectation for explosive chromospheric
evaporation. This is mainly because the solar flare studied in this
work occurred close to the solar limb (e.g., S26W66), and thus the
Doppler shifts are strongly affected by the projection effect.
Moreover, the chromospheric evaporation corresponds the second HXR
pulse, so the \fexxi line blueshifts could be seriously influenced
by the falling velocity of hot evaporating plasmas during the first
HXR pulse. We also derived time-distance images along the
IRIS slit (Figure~\ref{imag}g) from the CHASE data series
at \fei~6569.2~{\AA} and H$\alpha$~6562.8~{\AA}, as shown by the
overlaid green and cyan contours in panels~(a2) and (c2). It can be
seen that the \fei~6569.2~{\AA} emission source is co-spatial with
the blue-shifted velocity of \fexxi line at the flare kernel k1,
while the H$\alpha$~6562.8~{\AA} emission source is spatially
correlated with the red-shifted velocity of \siiv line at two flare
kernels. The feature of Doppler velocities and line widths suggests
an explosive chromospheric evaporation, namely the chromospheric
evaporation and condensation are simultaneously detected during the
C9.6 flare. The explosive chromospheric evaporation occurs nearly
simultaneously with the white-light enhancement.

\subsection{Solar jet}
\begin{figure}
\centering
\includegraphics[width=\linewidth,clip=]{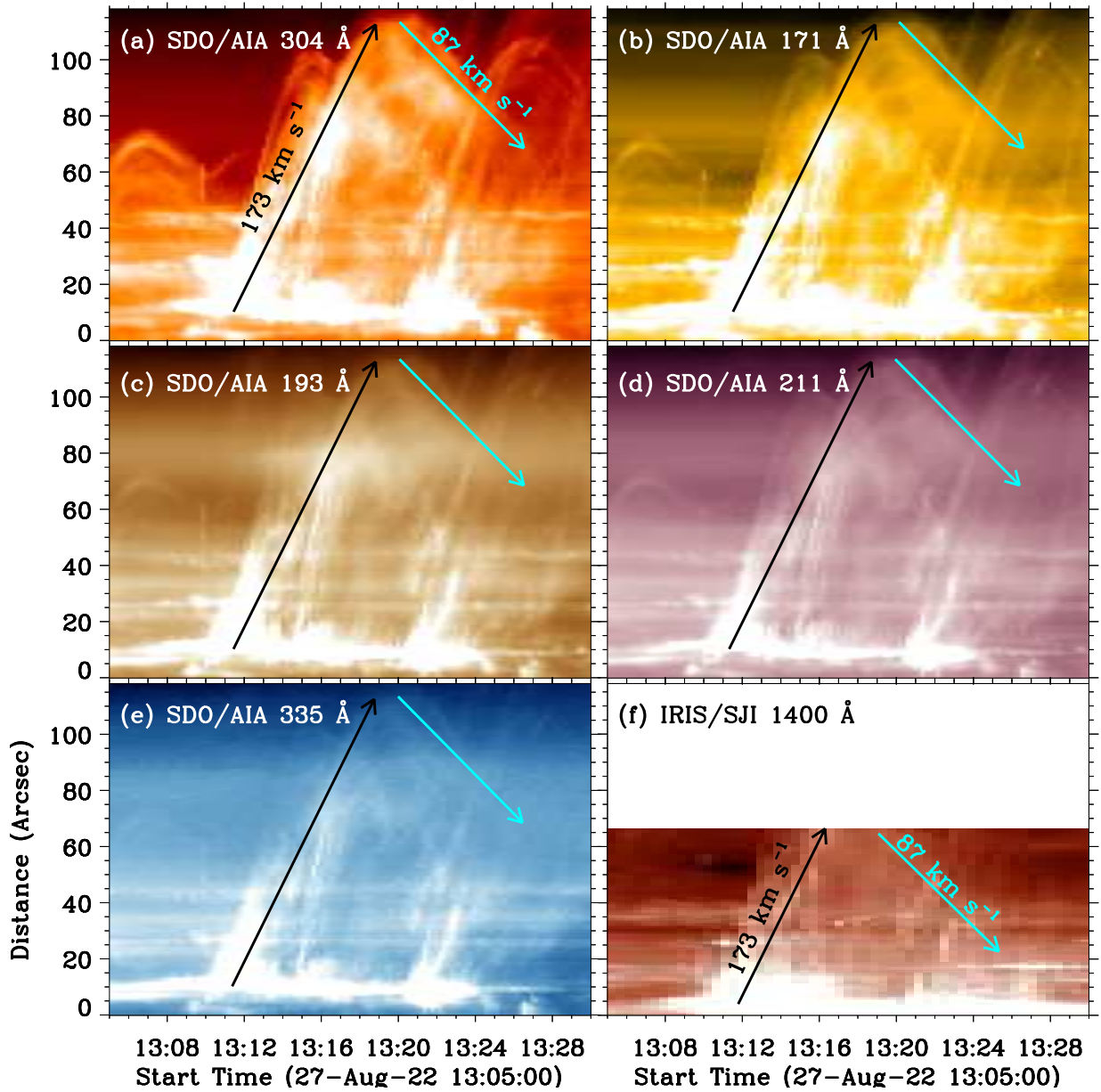}
\caption{Time-distance images along the solar jet, which are
performed from SDO and IRIS maps at wavelengths of AIA~304~{\AA}
(a), 171~{\AA} (b), 193~{\AA} (c), 211~{\AA} (d), and 335~{\AA} (e),
and SJI~1400~{\AA} (f). The black and cyan arrows indicate the rise
and fall speeds of the solar jet. \label{slice2}}
\end{figure}

A solar jet ejects from the flare loop-apex region, as shown in
Figure~\ref{imag}. To investigate its evolution, we draw the
time-distance images along the propagation direction of the jet, as
outlined by two green curves in Figure~\ref{imag}~(d). Here, a
constant width of $\sim$12\arcsec\ is used, so that the
bulk of the jet can be covered as much as possible during its whole
lifetime. Figure~\ref{slice2} presents the time-distance images at
wavelengths of AIA~304~{\AA}, 171~{\AA}, 193~{\AA}, 211~{\AA}, and
335~{\AA}, and SJI~1400~{\AA}. The solar jet firstly propagates
outward following after the C9.6 flare, and the apparent speed is
estimated to about 173~km~s$^{-1}$. Then some material of the solar
jet falls back with an apparent speed of about 87~km~s$^{-1}$. The
whole evolution of this jet could be seen in several AIA EUV
channels, suggesting that the solar jet is characterized by
multi-temperature plasma structures, which is similar to previous
results \citep[e.g.,][]{Lu19,Shen21}. IRIS/SJI did not observe the
whole solar jet, but similar speeds of propagating and falling are
detected, as shown in panel~(f). The dynamics of the jet
suggest that there is no mass ejecting from the Sun during the C9.6
flare, which is consistent to the absence of the matching CME in the
catalog. Therefore, the C9.6 flare is probable a confined flare.

\section{Discussion}
\begin{figure}
\centering
\includegraphics[width=\linewidth,clip=]{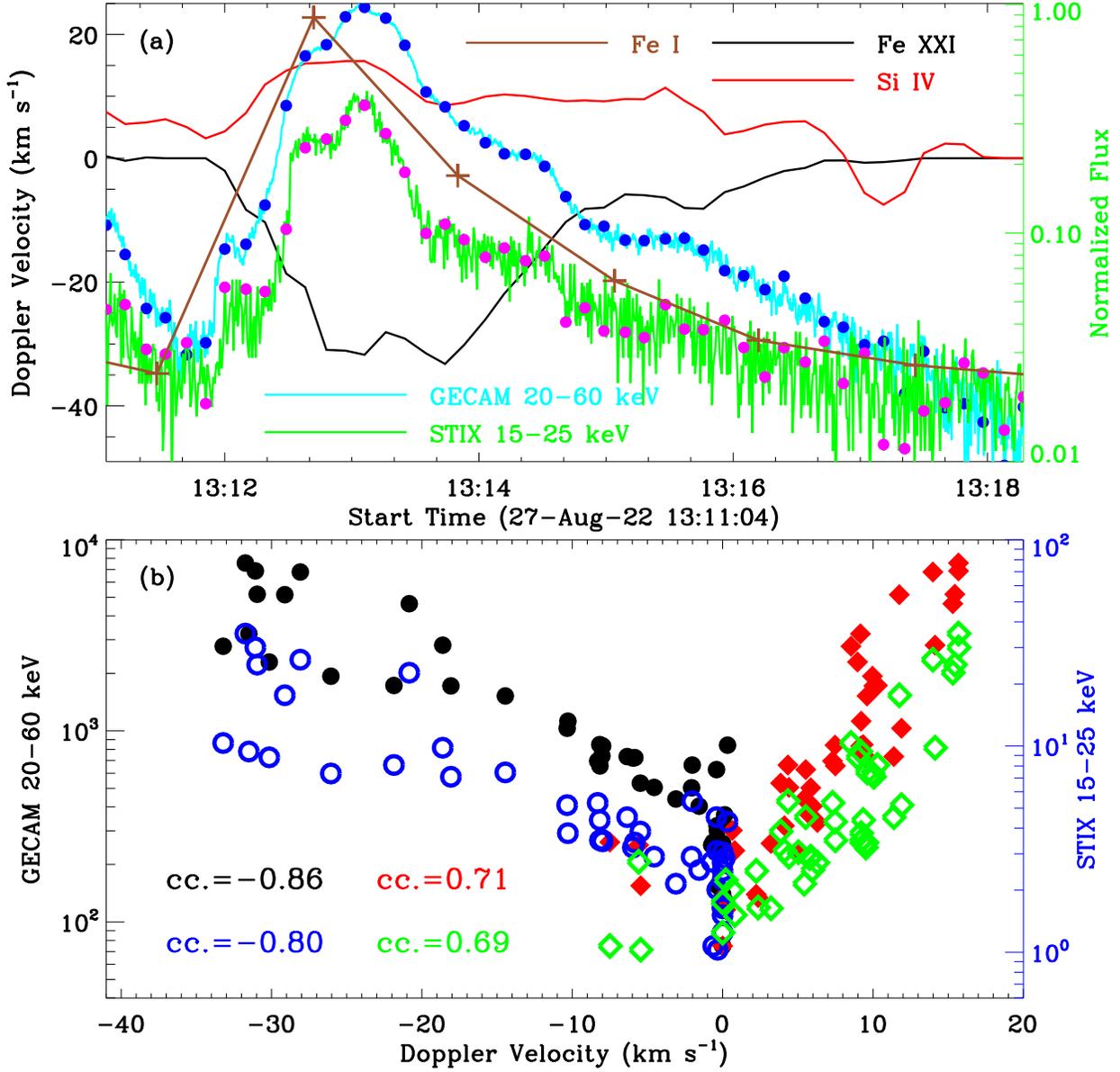}
\caption{Time series of the Doppler velocity in \fexxi\ (black) and
\siiv\ (red), and normalized HXR fluxes at GECAM~20$-$60~keV (cyan)
and STIX~15$-$25~keV (green), and the normalized white light flux at
CHASE \fei (brown). The blue and magenta dots mark these HXR points
used to perform the correlation analysis. Panel~(b): Scatter plots
of Doppler velocities of \fexxi\ (circles) and \siiv\ (diamonds)
depend on the HXR fluxes observed by GECAM (fill) and STIX (hollow).
Their Pearson correlation coefficients (cc.) are labeled.
\label{corr}}
\end{figure}

In this section, we discuss the physical mechanism of the explosive
chromospheric evaporation. Figure~\ref{corr} plots the correlation
between HXR emissions and Doppler velocities. Panel~(a) presents the
temporal evolution of Doppler velocities in \fexxi\ (black) and
\siiv (red), the normalized white-light flux at CHASE
\fei~6569.2~{\AA} (brown), and the normalized HXR fluxes at
STIX~15$-$25~keV (green) and GECAM~20$-$60~keV (cyan) during
13:11:04$-$13:18:14~UT, respectively. The flare line of \fexxi\
exhibits blueshifts, while the transition region line of \siiv\
reveals redshifts, and their maximum velocities are about
33~km~s$^{-1}$ and 16~km~s$^{-1}$, respectively. The
white-light flux seems to be temporally correlated with the Doppler
velocities of \fexxi\ and \siiv, and it is also consistent with HXR
fluxes recorded by GECAM and STIX, implying that both white-light
emission and chromospheric evaporation could be associated with
nonthermal electrons. The Doppler velocities and white-light flux
are averaged over the flare kernel (k1), as indicated by the two
short black lines in Figure~\ref{sp_iris2}. These curves appear to
show a good correlation with each other, but they have different
time cadences (Table~\ref{tab1}), making it impossible to correlate
them directly. Thus, we interpolate HXR fluxes to the IRIS time
cadence, as indicated by solid dots. It should be pointed out that
we do not interpolate the white-light flux, due to its very low time
resolution. We then show the dependence of Doppler velocities of
\fexxi\ (circles) and \siiv\ (diamonds) on HXR fluxes at
GECAM~20$-$60~keV (fill) and STIX~15$-$25~keV (hollow), as shown in
Figure~\ref{corr}~(b). High Pearson correlation coefficients are
found between the Doppler velocities of \fexxi\ and HXR emissions at
GECAM~20$-$60~keV (-0.86) and STIX~15$-$25~keV (-0.80), `$-$' means
that the Doppler velocities of \fexxi\ are blueshifts. The
Pearson correlation coefficients between red-shifted
velocities of \siiv\ and HXR fluxes are 0.71 and 0.69,
respectively. Such high correlation coefficients indicate
that the explosive chromospheric evaporation is probably driven by
nonthermal electrons accelerated by magnetic reconnection
\citep[e.g.,][]{Li15,Li17,Tian15}.

\begin{figure}
\centering
\includegraphics[width=\linewidth,clip=]{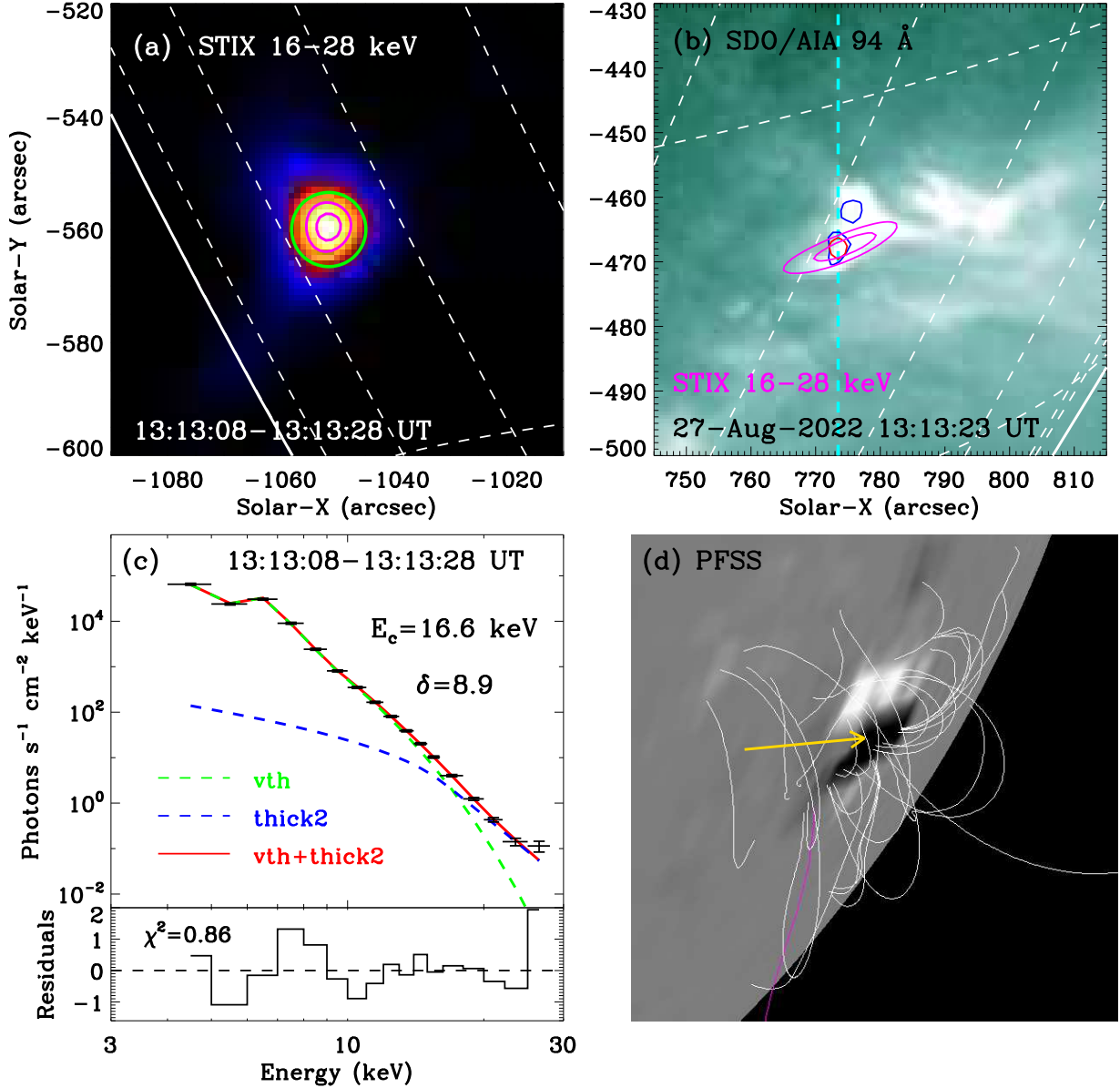}
\caption{Panels~(a) and (b): STIX~16$-$28~keV and AIA~94~{\AA} maps
during the solar flare. The magenta contours represent HXR emissions
at levels of 70\%, and 90\%. The dashed white lines indicate
latitude-longitude grids, and the solid white line marks the solar
limb. The green circle is an visibility forward-fit for the HXR
source area. The cyan dashed line outlines the IRIS slit, the blue
and red contours denotes to H$\alpha$ and \fei kernels. Panel~(c):
STIX X-ray spectrum between 4-28~keV with an integrated time of
20~s, the fitted thermal (green) and nonthermal (blue) components,
as well as the sum of both components (red). The low-energy cutoff
($E_c$), spectral index ($\delta$), and the chi-squared residual
($\chi^2$) are labeled. Panel~(d): The magnetic configuration from
the SDO viewpoint with a PFSS model. The white and purple lines
represent the closed and open magnetic field, respectively. The gold
arrow indicates the short flare loop. \label{stix_pfss}}
\end{figure}

To determine the energy flux that is deposited in the chromosphere
during the evaporation process, we need to constrain both the
electron flux and the footpoint area. We start by reconstructing the
nonthermal HXR source in the energy range of 16$-$28~keV with the
STIX pixelated science data using the Expectation Maximization
algorithm \citep{Massa19}. The resulting source is shown in
Figure~\ref{stix_pfss}~(a), with the contour levels set at 70\% and
90\% of the maximum intensity. It should be pointed out that Solar
Orbiter was located 147~degrees from the Earth\_Sun line, so in
contrast to the near-Earth observations, the flare is seen here near
the east limb. As was already evident from the observations in the
other wavelength range, this flare is very compact, and we
therefore cannot separate it into the expected double footpoint
structures. As the STIX source is rather circular, we performed an
additional image reconstruction using the Visibility Forward Fit
method \citep{Volpara22} with a circular source, as indicated by the
overplotted green circle. The advantage of this method is that it
returns the fit parameters with error estimates, and so we derive a
nonthermal source area ($A$) of
(4.1$\pm$0.7)$\times$10$^{17}$~cm$^2$. Since we most probably do not
resolve the source, this has to be considered as an upper limit.

While normally the precise HXR source locations are provided by the
STIX Aspect System \citep{Warmuth20b}, Solar Orbiter was too far
from the Sun (e.g., 0.79~AU) to get a reliable pointing solution. We
therefore reconstructed a thermal HXR image (6$-$10~keV) and
co-aligned it with a co-temporal EUV image at 174~{\AA} captured by
the Full Sun Imager (FSI) of the Extreme Ultraviolet Imager
\citep[EUI;][]{Rochus20} on Solar Orbiter.
Figure~\ref{stix_pfss}~(a) already shows the corrected source
location. In order to compare the nonthermal STIX source to the
other observations, we reproject the STIX image to the Earth view
assuming the source is located on the solar surface, which is
justified for a chromospheric footpoint source. This was performed
using version~4.1.5 of the SunPy open source software package
\citep{Sunpy20}. Figure~\ref{stix_pfss}~(b) shows the reprojected
STIX source overplotted on a co-temporal AIA~94~{\AA} image. The
apparent elongation of the source is a projection artefact due to
the fact that we do not resolve the east-west extension of the
source. However, we can see that the centroid of the reprojected
source appears to be coincident with the location of one H$\alpha$
kernel and the \fei source, as indicated by the magenta, blue, and
red contours, respectively. Thus, it could be regarded as one
footpoint of a short flare loop seen in AIA~94~{\AA}. Our
observation is consistent with previous findings, i.e., white-light
kernels are typically found to be co-spatial with HXR sources
\citep{Metcalf03,Cheng15,Yurchyshyn17}. Moreover, the IRIS slit
(dashed cyan line) crosses the H$\alpha$ kernel and HXR source,
confirming that the explosive evaporation is caused by nonthermal
electrons, which is similar to previous observations
\citep[e.g.,][]{Zhang16,Li18}.

In the next step, we constrain the energy input rate provided by the
injected nonthermal electrons by fitting the STIX count spectra in
the range of 4$-$28~keV with the combination of an isothermal
component and a thick-target power-law. This is performed with the
OSPEX spectral analysis
package\footnote{http://hesperia.gsfc.nasa.gov/ssw/packages/spex/doc/}
using the functions VTH and THICK2, with the addition of an albedo
component. Figure~\ref{stix_pfss}~(c) shows a fitted spectrum for a
representative time interval during the nonthermal HXR peak
(13:13:08$-$13:13:28~UT). In the plot, we indicate the low-energy
cutoff energy (E$_c$) and the spectral index ($\delta$) of the
injected electron flux. The $\chi^2$ value and the residuals plotted
in the lower part of the figure indicate a good fit result. Note
that the nonthermal spectrum is very soft. The cutoff energy of
16.6~keV confirms that the STIX X-ray radiation at 16$-$28~keV is
mainly from the nonthermal emission, thus the area ($A$) of
nonthermal source could be about
(4.1$\pm$0.7)$\times$10$^{17}$~cm$^2$. Next, the nonthermal power
($P$) above the cutoff energy could be calculated by the code in
SolarSoftWare, such as
``calc\_nontherm\_electron\_energy\_flux.pro", which is about
5.2$\times$10$^{27}$~erg~s$^{-1}$. At last, the nonthermal energy
flux ($P/A$) can be estimated to be about
(1.3$\pm$0.2)$\times$10$^{10}$~erg~s$^{-1}$~cm$^{-2}$. This is
just a bit larger than the critical threshold of deposited energy
flux given by \cite{Fisher85}. However, we have to consider our
result as a lower estimate only. We use the highest low-energy
cutoff that is consistent with the data, therefore our electron flux
is a lower estimate. Since the true low-energy cutoff is masked by
the thermal emission, the total electron flux could be considerably
higher. On the other hand, the HXR source is most probably not well
resolved, so the footpoint area could well be smaller. Since both
effects will result in a higher energy flux, we conclude that the
STIX observations support the explosive chromospheric evaporation
scenario.

In this flare, the HXR spectrum at higher energies is so
steep (i.e.,$\delta$=8.9) that it may actually be consistent with a
hotter secondary thermal component as opposed to injected electrons
with a soft spectrum. To investigate this possibility, we also
fitted the STIX spectrum that showed a high-energy tail with
double-thermal components. For the time interval shown in
Fig.~\ref{stix_pfss}~(c), this resulted in a cooler component with a
temperature ($T$) of $(16.4\pm0.5$)~MK and an emission measure
($EM$) of ($2.8\pm0.3)\times10^{48}$~cm$^{-3}$, and a hotter
component with $T=(31.9\pm4.3)$~MK and $EM=(1.7\pm1.7)\times
10^{46}$~cm$^{-3}$. It is evident that the hotter component is
significantly less well constrained than the cool one. However, the
strongest argument against the interpretation of the high-energy
tail as a hotter thermal component is the cooling time, which should
be as short $\approx$2~minutes in order to explain the quickly
vanishing high-energy tail in the observed spectrum. To estimate the
cooling time of the hotter component, we use the model of
\citet{Cargill95}. This is based on conductive and radiative cooling
timescales assuming an isotropic, isothermal, collisionless plasma.
Apart from temperature, the model requires the electron density and
the loop half-length as input. We get a density of
$\sim$$9\times10^{9}$~cm$^{-3}$ from the EM of the hotter component
and the source volume, which we derive from the forward-fitted
circular source assuming spherical geometry. Taking the footpoints
of the loop structure seen in the AIA~94~{\AA} image and assuming a
semicircular loop, we obtain a loop half-length of about 7~Mm. For
these parameters, the Cargill model predicts a time of
$\sim$11~minutes for the plasma to cool from $\sim$32~MK down to
$\sim$16~MK, which is significantly longer than the observed decay
time of the high-energy tail of the spectrum. Even when we assume a
very small filling factor of just 0.01 for the hotter component
(which results in higher densities and thus more efficient cooling),
the predicted cooling time is still $\sim$7~minutes. In summary, the
observations are indeed more consistent with a nonthermal
thick-target model.

In order to look closely at the magnetic field topology on the source
surface that connects to the flare region, we perform a potential
field source surface (PFSS) extrapolation to the synoptic SDO/HMI
map, as shown in Figure~\ref{stix_pfss}~(d). We notice that the
flare area is dominated by closed magnetic field lines (white
lines), preventing electron beams generated from the C9.6 flare to
escape into the interplanetary space. Therefore, we can not
observe the radio III burst at lower frequencies, as shown in
Figure~\ref{flux}~(a). So, the C9.6 flare is confined rather than
eruptive. Moreover, some closed magnetic lines appear at the region
of the short flare loop seen in AIA~94~{\AA}, as indicated by the
gold arrow. This further suggests the presence of flare loop that
connects double H$\alpha$ kernels.

\section{Summary}
Based on multi-wavelength observations measured by CHASE, IRIS,
SDO/AIA, SDO/HMI, STIX, GECAM, and SWAVES, we investigate explosive
chromospheric evaporation during a C9.6 flare on 2022 August 27. The
main conclusions are summarized as following:

\begin{enumerate}

\item The H$\alpha$ spectroscopic observations obtained with CHASE reveal
two flare kernels, which are connected by a short flare loop seen in
the high-temperature channel of AIA (131~{\AA} and 94~{\AA}). Two
loop legs are rooted in positive and negative magnetic fields,
respectively. Thus, they are recognized as conjugated footpoints.
The double kernels (k1 and k2) show red-shifted velocities of less
than 10~km~s$^{-1}$ in H$\alpha$ Doppergrams. One kernel (k1)
exhibits red shifts of 10$-$20~km~s$^{-1}$ in the cool lines of \ci\
and \siiv\ detected by IRIS. Those red shifts can be regarded as the
downflows driven by chromospheric condensation
\citep{Li04,Zhang16,Libbrecht19,Graham20,Li22}.

\item One flare kernel (k1) shows blue shifts of 30$-$40~km~s$^{-1}$ in
the hot line of \fexxi. The blueshift could be the indicator of the
upflow caused by chromospheric evaporation
\citep{Brosius15,Dudik16}. Interestingly, the blue-shifted velocity
is co-spatial with the white-light emission captured by CHASE at
\fei~6569.2~{\AA}, implying that the chromospheric evaporation and
white-light emission have the same driver. We also notice that the
velocity of upflows is rather small compared to previous findings
\citep{Veronig10,Li15,Young15,Tian18}, which could be due to
the fact that the WLF was located at the solar limb, or because that
the chromospheric evaporation spans from impulsive phase to the
gradual phase of the C9.6 flare.

\item The chromospheric condensation and evaporation are simultaneously
observed during the C9.6 flare, implying an explosive chromospheric
evaporation. A lower boundary for the nonthermal energy flux is
estimated to be
(1.3$\pm$0.2)$\times$10$^{10}$~erg~s$^{-1}$~cm$^{-2}$, confirming
the explosive chromospheric evaporation
\citep[e.g.,][]{Fisher85,Li18,Sadykov19}. We also considered
alternative fittings to the STIX spectrum with double-thermal
components. However, the cooling time is too long for the hotter
plasma that is required.

\item High correlations are found between Doppler velocities of \fexxi\
and \siiv\ and the HXR emission at GECAM~20$-$60~keV and
STIX~15$-$25~keV. Moreover, the nonthermal deposited energy source
seen at STIX~16$-$28~keV is coincident with one flare kernel seen in
H$\alpha$ and \fei wavebands. All those observational facts indicate that
the explosive chromospheric evaporation is driven by nonthermal
electrons that accelerated by the magnetic reconnection
\citep[e.g.,][]{Tian15,Warren16,Zhang16,Lee17,Li17}.

\item The CHASE map at \fei~6569.2~{\AA} shows a white-light enhancement
at one flare kernel (k1), and the similar enhancement is also
detected in the pseudo-intensity map derived from SDO/HMI continuum
filtergrams \citep[cf.][]{Song18}. Thus, the C9.6 flare is indeed a
white-light flare. The white-light flux at \fei~6569.2~{\AA} appears
to be temporally correlated with the HXR flux, indicating that the
white-light emission is induced by nonthermal electrons. PFSS
extrapolation based on the synoptic SDO/HMI map suggests that the
flare area is characterized by closed magnetic field lines. No
flare-related lower-frequency radio III burst or CMEs are observed,
and a jet originated from the loop-apex finally falls back,
suggesting that the white-light flare is also a confined flare
\citep[e.g.,][]{Song14,Kliem21}.
\end{enumerate}

\acknowledgments The authors would like to thank the referee
for his/her inspiring comments and valuable suggestions. We also
thank the teams of CHASE, IRIS, SDO, STIX, GECAM, and GOES for their
open data use policy. This work is funded by the National Key R\&D
Program of China 2022YFF0503002 (2022YFF0503000), the NSFC under
grant 11973092, 12073081. D.~L. is also supported by the Surface
Project of Jiangsu Province (BK20211402) and the Specialized
Research Fund for State Key Laboratories. F.~S. is supported by the
German Space Agency (DLR) grant No. 50 OT 1904. The CHASE mission is
supported by China National Space. The CHASE mission is supported by
China National Space Administration (CNSA). Solar Orbiter is a space
mission of international collaboration between ESA and NASA,
operated by ESA. The STIX instrument is an international
collaboration between Switzerland, Poland, France, Czech Republic,
Germany, Austria, Ireland, and Italy.

\begin{table}
\caption{Details of observational instruments used in this paper.}
\label{tab1}
\centering
\tabcolsep 3pt
\begin{tabular}{c c c c c c c c}
\toprule[1.2pt]
Instruments &  Windows  & Wavelengths            & Spectral  Dispersion & Cadence  &  Spatial resolution$^1$    \\
\midrule[0.8pt]
           &  H$\alpha$ & 6559.7$-$6565.9~{\AA}  &   $\sim$48~m{\AA} &  $\sim$71~s   &  2.08\arcsec    \\
CHASE      &  \fei\    &  6567.8$-$6570.6~{\AA}  &   $\sim$48~m{\AA} &  $\sim$71~s   &  2.08\arcsec    \\
\midrule[0.8pt]
            &  \oi\    &  1351.42$-$1356.61~{\AA}  &     25.96~m{\AA}     & $\sim$9.4~s  &  0.33\arcsec    \\
IRIS/SP     &  \siiv\  &  1398.66$-$1407.00~{\AA}  &     25.44~m{\AA}     & $\sim$9.4~s  &  0.33\arcsec    \\
IRIS/SJI    &   FUV    &   1400~{\AA}              &      --              & $\sim$28.2~s &  0.33\arcsec    \\
\midrule[0.8pt]
            &   EUV    &     94$-$335~{\AA}        &      --              &    12~s      &    1.2\arcsec   \\
SDO/AIA     &   UV     &     1600~{\AA}            &      --              &    24~s      &    1.2\arcsec   \\
            & Magnetogram &   6173~{\AA}           &      --              &    45~s      &    1.2\arcsec   \\
SDO/HMI     &  Continuum  &   6173~{\AA}           &      --              &    45~s      &    1.2\arcsec   \\
\midrule[0.8pt]
            & SXR      &    1$-$8~{\AA}            &      --              &    1~s        &       --       \\
GOES        & SXR      &    0.5$-$4~{\AA}          &      --              &    1~s        &       --       \\
\midrule[0.8pt]
            &  LC   &     4$-$10~keV               &      --              &    0.5~s      &       --       \\
            &  LC   &     10$-$15~keV              &      --              &    0.5~s      &       --       \\
STIX        &  LC   &     15$-$25~keV              &      --              &    0.5~s      &       --       \\
     &  Spectra     &     4$-$28~keV               &    $\sim$1~keV       &    20~s       &       --       \\
     &  Pixel       &     16$-$28~keV              &      --              &    20~s       &    11.5\arcsec  \\
\midrule[0.8pt]
GECAM       & HXR   &    20$-$60~keV               &      --              &    0.2~s      &        --       \\
\midrule[0.8pt]
SWAVES      & Radio &   0.023$-$16.025~MHz         &      --              &    60~s       &        --       \\
\bottomrule[1.2pt]
\end{tabular}
\begin{tablenotes}
   \item $^1$The spatial resolution here is seen from Earth, i.e., 1~AU.
\end{tablenotes}
\end{table}

\end{document}